# Investigation of Six Imidazolium-Based Ionic Liquids as Thermo-Kinetic Inhibitors for Methane Hydrate by Molecular Dynamics Simulation

Mohammad Ebrahim Haji Nasrollah[a], Bagher Abareshi[a], Cyrus Ghotbi[*a],
Vahid Taghikhani[a], Amir Hossein Jalili[b],

[a] *Department of Chemical and Petroleum Engineering, Sharif University of Technology, Azadi Ave., Tehran, Iran*

*Corresponding Author's E-mail:   cghosfd@yahoo.com

**Abstract**
The thermo-kinetic inhibition mechanism of six imidazolium-based ionic liquids (ILs) on methane clathrate hydrate formation and growth is studied in this work using classical molecular dynamics (MD) simulation. The ionic liquids investigated include 1-(2,3-dihydroxypropyl)-3-methylimidazoliumbis(fluorosulfonyl)imide ($[C_3(OH)_2mim][f_2N]$), 1-(2-hydroxyethyl)-3-methylimidazolium bis(fluorosulfonyl)imide ($[C_2OHmim][f_2N]$), 1-ethyl-3-methylimidazolium tetrafluoroborate ($[C_2mim][BF_4]$), 1-butyl-3-methylimidazolium tetrafluoroborate ($[C_4mim][BF_4]$), 1-butyl-3-methylimidazolium acetate ($[C_4mim][OAc]$) and 1-ethyl-3-methylimidazolium ethylsulfate ($[C_2mim][EtSO_4]$). Simulations showed that $[C_2OHmim][f_2N]$ and $[C_3(OH)_2mim][f_2N]$ are strongly hydrated compared to other ILs because of hydrogen bonding between OH groups of the cation and water molecules. They also exhibit high diffusion rates towards crystal surface and bond to it through strong intermolecular interactions. As a result, these two ILs are stronger thermo-kinetic inhibitors for formation and growth of methane hydrates compared to other ILs studied in this work as well as conventional inhibitors like methanol and NaCl. The simulations also revealed that cations of $[C_3(OH)_2mim][f_2N]$ and $[C_2OHmim][f_2N]$ show that the presence of ions near the hydrate crystal causes hindrance for water and guest molecules adsorbing on the hydrate surface, which inhibits the growth of hydrate crystals. In addition, it is shown that $[C_3(OH)_2mim][f_2N]$ and $[C_2OHmim][f_2N]$ are more likely to inhibit hydrate formation.

**Keywords: methane hydrate, molecular dynamics simulation, ionic liquid, kinetic inhibitor, thermodynamic inhibitor.**

**Research Highlights:**
- Investigation of kinetic inhibition mechanism of six imidazolium-based ILs on methane hydrates formation and growth using MD simulations.
- Investigation of thermodynamic inhibition mechanism of six imidazolium-based ILs on methane hydrates growth by MD simulations.
- Drawing a comparison between methane hydrate inhibition Effectiveness of ILs.

*1. Introduction:*

Methane hydrates, which could be considered as a source of energy [1], are commonly formed during natural gas production operations under certain temperature and pressure conditions. Formation of stable gas hydrates in gas and oil production and pipelines can lead to severe safety problems and huge economic loss [2-3]. Therefore, several different methods have been developed in order to retard gas hydrate formation. Thus, hydrate inhibitors have achieved widespread popularity. There are two types of inhibitors, namely thermodynamic and kinetic inhibitors. Most of the free water is hydrogen bonded to the inhibitor molecules by injecting thermodynamic inhibitors. This reduces the water activity so that lower temperatures and higher pressures are required to form hydrates at low concentrations of non-hydrogen bonded water [4]. Kinetic inhibitors bond to the hydrate surface and prevent plug formation for a period longer than the free water residence time in a pipeline. It has also been found that a combination of thermodynamic and kinetic inhibitors is still required to give better results in some cases [5]. Thus, novel and more effective inhibitors are in demand.

For this purpose, various types of ionic liquids (ILs) have been studied [6-10]. Ionic liquids have strong electrostatic charges and at the same time their anions and/or cations can be chosen or tailored to form hydrogen bonding with water [9]. A series of experiments have been carried out to investigate the use of ILs as novel hydrate inhibitors and the discovery of dual function thermo-kinetic inhibitors [6-8].The problem is that, testing the ability of ILs as thermo-kinetic inhibitors involves costly experiments. On the other hand, computers are becoming cheaper and more powerful than ever. Therefore, it is reasonable to characterize the effects of ILs as inhibitors on methane hydrate before performing expensive, long-running experiments. Few works have been performed in order to investigate the effects of some common kinetic inhibitors [10-12]. Nevertheless, to the best of our knowledge; the thermo-kinetic inhibiting functions of ILs have not yet been studied by MD simulations.

In this paper, thermodynamic and kinetic effects of five dialkylimidazolium-based ILs on formation and growth of methane hydrate crystals have been studied by MD simulation. Then, the performance of these ILs as thermo-kinetic inhibitors were investigated and finally comparisons were made between different ILs to identify the perfectly suitable inhibitors. In addition, some comparisons were made between ILs and other common inhibitors such as methanol in order to have a criterion to measure thermodynamic effectiveness of ILs in comparison with common inhibitors.

*2. Computational method and simulation details*

The ILs studied in this work are listed in Table 1. As was mentioned the ILs are comprised of an imidazolium cation with two substituents attached to the nitrogen atoms. One of the substituents is a methyl group and the other one, –R is either a linear alkyl chain or a hydroxyl substituted alkyl chain.

**Table 1-Ionic Liquids Studied in this Work**

| Symbol | Chemical name | –R substituent | Anion |
|---|---|---|---|
| [$C_2$OHmim][$f_2$N] | 1-(2-hydroxyethyl)-3-methylimidazolium bis(fluorosulfonyl)imide | -$CH_2$-$CH_2$-OH | [$N(SO_2F)_2$]$^-$ |
| [$C_3(OH)_2$mim][$f_2$N] | 1-(2,3-dihydroxypropyl)- 3-methylimidazoliumbis(fluorosulfonyl)imide | -$CH_2$-CH(OH)-$CH_2$-OH | [$N(SO_2F)_2$]$^-$ |
| [$C_2$mim][$BF_4$] | 1-ethyl-3-methyl imidazoliumtetrafluoroborate | -$CH_2$-$CH_3$ | [$BF_4$]$^-$ |
| [$C_4$mim][$BF_4$] | 1-butyl-3-methylimidazolium tetrafluoroborate | -$CH_2$-$CH_2$-$CH_2$-$CH_3$ | [$BF_4$]$^-$ |
| [$C_4$mim][OAc] | 1-butyl-3-methylimidazolium acetate | -$CH_2$-$CH_2$-$CH_2$-$CH_3$ | [$CH_3COO$]$^-$ |
| [$C_2$mim][$EtSO_4$] | 1-ethyl-3-methylimidazoliumethylsulfate | -$CH_2$-$CH_3$ | [$CH_3$-$CH_2$-$OSO_3$]$^-$ |

In this study, all the simulations were carried out using TINKER software version 03 package [13]. In order to prepare the ILs for MD simulations, suitable force fields and potential functions must be chosen. For [$C_2$mim][$BF_4$] and [$C_4$mim][$BF_4$] a model based on the AMBER force field [14] with modifications made by Liu et al. [15] was selected. In order to prepare [$C_2$OHmim][$f_2$N], [$C_3(OH)_2$mim][$f_2$N], and [$C_2$mim][$EtSO_4$] for simulations, models based on the OPLS-AA [16] potential function and parameters presented by Lopez et al. [17-21] were selected. The OPLS-AA potential parameters presented by Bowron et al. [22] was used to simulate [$C_4$mim][OAc].

The TIP4P/ice force field, which has been presented by Abascal et al. [23], and used in other works research [24] was selected to describe water molecules in this study. In addition, OPLS-AA force field was used to describe methane molecules. Two types of simulation runs were designed in order to investigate the effects of ILs on methane hydrates; one for studying the growth of methane hydrates and the other one for studying the hydrate formation. In the former case, water molecules as the liquid phase were exposed to a hydrate crystal as the solid phase. The hydrate crystal was made of 2×2×2 unit cells of structure I hydrate. Then, molecules of IL, corresponding to about 0.5 and 1 mol% of IL, were added to the liquid phase in the simulation box. Afterwards, simulation started at 250 K and 5 atm for 3 ns. In the second box, water and methane molecules (90 mol% water and 10 mol% methane) were arranged in the simulation box and then molecules of IL (about 1 mol%) were added. The simulations were performed in the *NPT* ensemble at 200 K and 20 atm for 35 ns. Because of time limitations, the simulations were performed in harsh temperature and pressure conditions to identify hydrate formation faster in this case. The dimensions of the simulation boxes were designed 50Å×24Å×24Å. The primary simulation boxes for each case before adding ILs are shown in Fig. 1. In each case, simulations with blank samples, i.e. without inhibitor, were also performed.

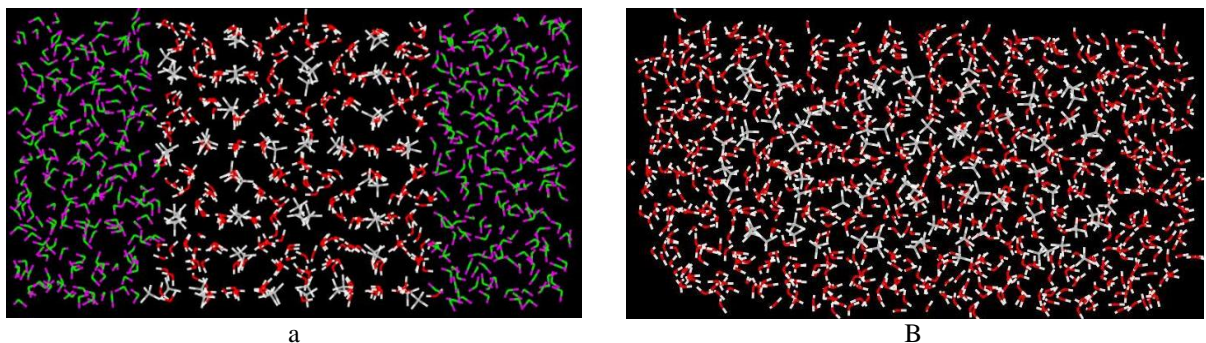
a          B

**Figure 1-Simulation box containing (a) methane hydrate crystal at the center of the box surrounded by water molecules, (b) mixture of methane and water molecules.**

## 3. Results and Discussion

### 3.1 Thermodynamic inhibition

Normally a phase transformation is considered relative to the change in Gibbs free energy. Upon addition of an inhibitor, the Gibbs free energy is increased. As we know, the value of $\Delta H$ is relatively constant [4]. Therefore, it appears that the energetic effects are not appreciably affected by the inhibitors. In order to increase the Gibbs free energy, the inhibitor primarily affects the structure of the water phase, i.e. it encourages non randomness or structures other than hydrate-like clusters in water. Therefore, the mechanism of formation inhibition is aided by increased competition for water molecules by the dissolved inhibitor molecule or ion through hydrogen bonding for alcohols or glycols, or via Columbic forces (for ions). Consequently, the more the IL affects the structure of the water phase, the better thermodynamic inhibitor is the IL.

In dilute aqueous ionic liquids most of the ions are isolated within bulk water [25]. In this part, we focus on the effects of ILs on the surrounding water structure through examining the radial distribution functions (RDFs). It should be noted that at this stage, we used the results from the simulation box, which was prepared to study the methane hydrate growth.

In order to study the thermodynamic inhibition effectiveness of ILs it is better to compare RDFs between oxygen atoms on water molecules ($g_{o-o}$) in different systems. Simulation without any inhibitor (blank system) was performed to provide a good criterion in which we can compare RDFs from inhibited systems with RDFs from blank system. Better thermodynamic inhibitors cause more disorder in the structure of water molecules in liquid phase. RDFs with intense maximums at shorter distances (compared to blank system) represent more disorder in the water phase. $g_{o-o}$ for systems containing ILs and blank system are shown in Fig.2.

Systems containing $[C_2mim][BF_4]$, $[C_4mim][BF_4]$, $[C_4mim][OAc]$ and $[C_2mim][EtSO_4]$ have RDFs similar to blank system. However, systems containing $[C_3(OH)_2mim][f_2N]$ and $[C_2OHmim][f_2N]$ highly affect the water phase. Thus, these two ILs are expected to be better thermodynamic inhibitors. In general, this figure shows small, but detectable differences between the inhibited systems. It can be realized that $[C_3(OH)_2mim][f_2N]$ and

[C$_2$OHmim][f$_2$N] ILs cause stronger clustering effect on water molecules in comparison with other ionic liquids, which can be ascribed to their ability to form strong hydrogen bonds with water molecules through OH functional groups in [C$_3$(OH)$_2$mim]$^+$ and [C$_2$OHmim]$^+$. On the other hand, [C$_2$mim]$^+$ and [C$_4$mim]$^+$ are not able to form hydrogen bonds, thus they show the least hydration.

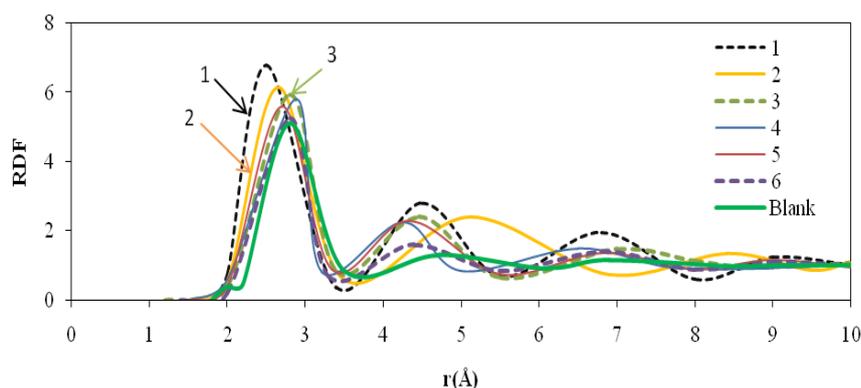

**Figure 2-** RDFs between oxygen atoms on water molecules (g$_{O-O}$) for systems containing different ILs; (1) [C$_3$(OH)$_2$mim][f$_2$N] , (2) [C$_2$OHmim][f$_2$N] , (3) [C$_2$mim][BF$_4$] , (4) [C$_4$mim][BF$_4$] , (5) [C$_2$mim][OAc] , (6) [C$_2$mim][EtSO$_4$] and (7) Blank system.

Therefore, the thermodynamic inhibition effectiveness of these ILs is in the following order:

[C$_3$(OH)$_2$mim][f$_2$N]>[C$_2$OHmim][f$_2$N]>[C$_2$mim][BF$_4$]>[C$_4$mim][BF$_4$]>[C$_4$mim][OAc]>[C$_2$mim][EtSO$_4$]

In order to make a comparison between [C$_3$(OH)$_2$mim][f$_2$N] and [C$_2$OHmim][f$_2$N] ILs and common thermodynamic inhibitors, similar simulations with 10 wt% NaCl and 10 wt% methanol were carried out and g$_{o-o}$ for each system was calculated (Fig.3).

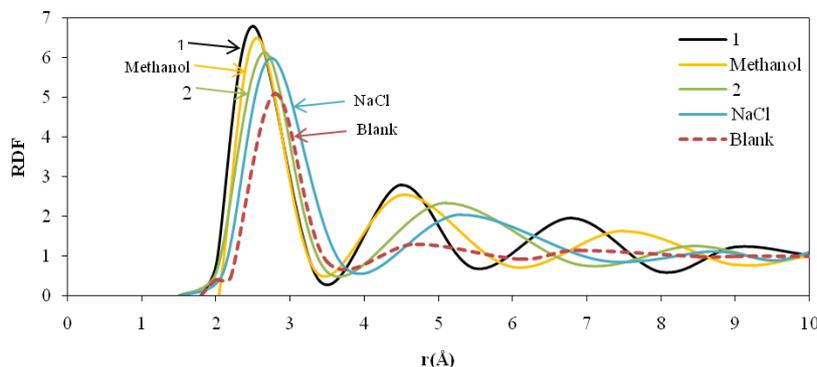

**Figure 3-** RDFs of oxygen atoms on water molecules (g$_{O-O}$) for (1) [C$_3$(OH)$_2$mim][f$_2$N], (2) [C$_2$OHmim][f$_2$N], Methanol, NaCl and Blank systems. Comparison between [C$_3$(OH)$_2$mim][f$_2$N], [C$_2$OHmim][f$_2$N] ILs and NaCl, methanol with 10 wt% as common thermodynamic inhibitors were carried out and g$_{o-o}$ for each system was calculated.

It can be observed that systems containing [C$_3$(OH)$_2$mim][f$_2$N], [C$_2$OHmim][f$_2$N] and methanol have approximately similar disorders in their structure of water phase. Thus, one can postulate that methanol and these two ILs have similar thermodynamic inhibition effectiveness. Having said that, [C$_3$(OH)$_2$mim][f$_2$N] seems to be a better thermodynamic inhibitor because this IL changed the structure of liquid water more than others. Methanol and [C$_2$OHmim][f$_2$N] are the best inhibitors after [C$_3$(OH)$_2$mim][f$_2$N].

## *3.2 Kinetic inhibition*

We investigate the kinetic inhibition effectiveness of ILs in two parts. At first, we analyze the results obtained from the first simulation runs (kinetic inhibition of methane hydrate growth) and then we study the second simulation run results (kinetic inhibition of methane hydrate formation).

### 3.2.1 Kinetic inhibition of methane hydrate growth

As mentioned, kinetic inhibition mechanism is to prevent water molecules from getting close to the crystal by bonding to the crystal surface. Therefore we need to know which ion moves faster towards the hydrate crystal. This movement could be a criterion for attractive interaction between the cation or anion of IL and the hydrate crystal, indicating better attachment and hence vigorous inhibition effect.

For this purpose mobility of different ions was calculated by molecular dynamics simulation (Table 2). It should be mentioned that Einstein relation, Eq. 1, was used to compute mobility of these cations at 250 K.

$$mobility = \frac{1}{6} \lim_{t \to \infty} \frac{d}{dt} \langle \Delta r(t)^2 \rangle \quad (1)$$

where $\langle \Delta r(t)^2 \rangle$ is the mean square displacement (MSD) of the center-of-mass of the molecule or ion.

Table 2- Mobility of different ions calculated by molecular dynamics simulation.

| Cation | mobility (cm$^2$s$^{-1}$) × 10$^7$ | Anion | mobility (cm$^2$s$^{-1}$) × 10$^7$ |
|---|---|---|---|
| [C$_3$(OH)$_2$mim]$^+$ | 24.807 | [f$_2$N]$^-$ | 11.192 |
| [C$_2$OHmim]$^+$ | 22.196 | [OAc]$^-$ | 8.492 |
| [C$_2$mim]$^+$ | 18.473 | [EtSO$_4$]$^-$ | 6.817 |
| [C$_4$mim]$^+$ | 15.267 | [BF$_4$]$^-$ | 5.018 |

Simulation results show cations generally reveal higher mobility rather than anions. Since higher mobility reveals more attraction force between ions and the hydrate crystal, it can be said that the effect of cation type on the kinetic inhibition effectiveness is more significant than that of anion type. In addition, Table 2 shows that the attraction forces between hydrate crystal and $[C_3(OH)_2mim]^+$ is more than other ions so it can have the best kinetic inhibition.

Up until now it is understood that in all systems, when simulation started, cations tended to move towards the hydrate crystal and positioned themselves at hydrate-liquid interface. These cations are likely to bond to hydrate surface and prevent water molecules from getting close to hydrate surface. Now we have to investigate how much the ionic liquids prevent water molecules from getting close to hydrate surface. For this purpose, the local density of water molecules, which could be a criterion for accumulation of water molecules at the hydrate-liquid water interface, has been defined as below:

$$\text{Local density of water molecules} = \frac{\dfrac{\text{Number of water molecules in the element}}{\text{Volume element}}}{\dfrac{\text{Number of water molecules in the box}}{\text{total volume}}}$$

Local density of water molecules can be calculated in different periods of time to study the kinetic inhibition effectiveness of ILs. Therefore the simulation box was divided into similar volume elements with the dimensions of 1Å×24Å×24Å. Then local density values in the volume element next to the hydrate crystal were calculated for different periods of time for systems containing 0.5 mol% and 1 mol% of ILs. The local density of water molecules during the simulation is shown in Fig. 4.

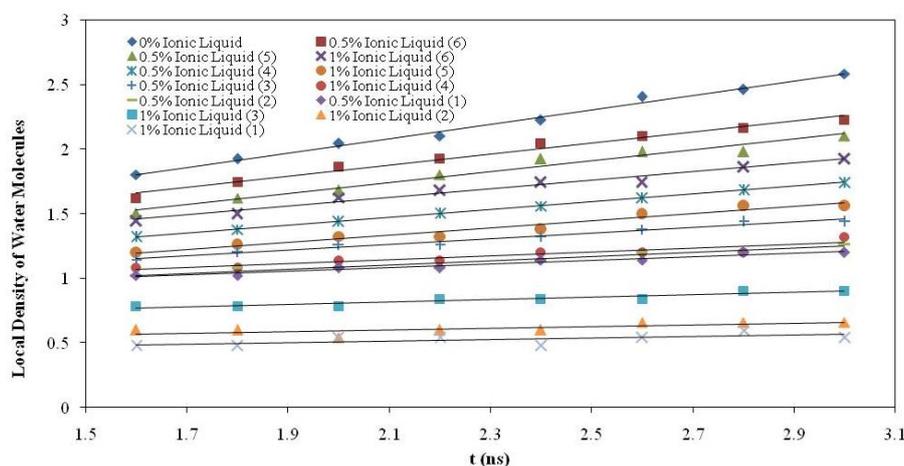

**Figure 4-Local density of water molecules at hydrate crystal-liquid interface, (1) $[C_3(OH)_2mim][f_2N]$, (2) $[C_2OHmim][f_2N]$, (3) $[C_2mim][BF_4]$, (4) $[C_4mim][BF_4]$, (5) $[C_4mim][OAc]$ and (6) $[C_2mim][EtSO_4]$.**

As time passes, water molecules are adsorbed to the hydrate crystal. Consequently local density of water molecules increases as a function of time. The slope of local density line in the blank system is 0.558, but for systems containing IL, this slope has lower values. Table 3 presents the slopes of local density lines in different systems. It can be realized that kinetic inhibition effectiveness of [C$_3$(OH)$_2$mim][f$_2$N] is more than that of other ILs because at any concentration: (1) the slope of local density line in the system containing [C$_3$(OH)$_2$mim][f$_2$N] is less than others and (2) the values of local density of water molecules are the least in comparison with other ILs at any time.

Thus, we can list the kinetic inhibition effectiveness of ILs in the following order:

[C$_3$(OH)$_2$mim][f$_2$N]>[C$_2$OHmim][f$_2$N]>[C$_2$mim][BF$_4$]>[C$_4$mim][BF$_4$]>[C$_4$mim][OAc]>[C$_2$mim][EtSO$_4$]

Table 3-The slope of local density line in systems containing 0.5 and 1 mol% of ionic liquid.

| Ionic Liquid | 0.5 mol% of ionic liquid | 1 mol% of ionic liquid |
|---|---|---|
| [C$_3$(OH)$_2$mim][f$_2$N] | 0.143 | 0.057 |
| [C$_2$OHmim][f$_2$N] | 0.161 | 0.064 |
| [C$_2$mim][BF$_4$] | 0.221 | 0.096 |
| [C$_4$mim][BF$_4$] | 0.300 | 0.150 |
| [C$_4$mim][OAc] | 0.418 | 0.275 |
| [C$_2$mim][EtSO$_4$] | 0.425 | 0.332 |

*3.2.2 Kinetic inhibition of methane hydrate formation*

In this section, results obtained from the second simulation box were analyzed in order to study the kinetic inhibition effectiveness of different ILs in delaying methane hydrate formation. Seven simulation runs were prepared, one of which was a blank system and the others contained about 1 mol% IL. All the simulations were carried out at 200 K and 20 atm for 35 ns. As it is shown in Fig. 5a, the RDF between carbon atoms in methane hydrate crystal has two maximums in 6.5 Å and 10.5 Å. The RDF between carbon atoms in the blank system before running the simulations was also calculated which has maximums in 4.2 Å, 6.2 Å and 8.2Å (Fig. 5b).

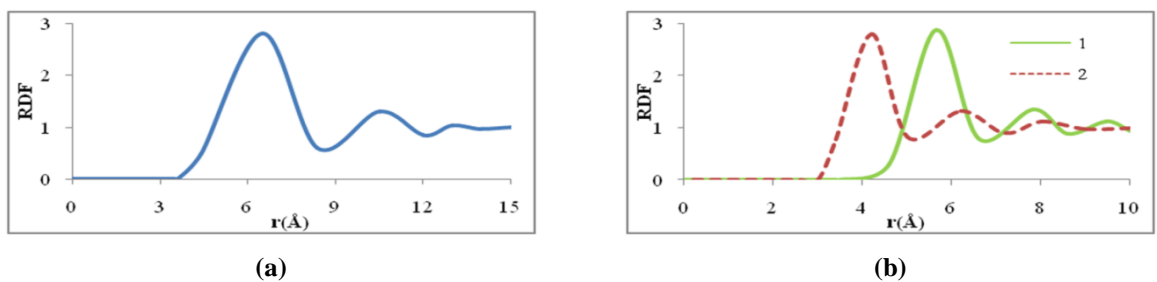

(a)      (b)

**Figure 5-(a) RDF between carbon atoms in methane hydrate crystal and (b) RDF between carbon atoms in the blank system (1) at the beginning and (2) after 35 ns.**

Maximums in blank system are expected to be shifted and move to the right as time passes because of tendency of the system to form methane hydrate at this temperature and pressure. After 35 ns, the maximums of blank system appeared in 6.12 Å and 8.47 Å (Fig. 5b). In other systems containing different ILs, the maximums shifted to the right less than the blank system, which means that ILs had kinetic inhibition effects. A good kinetic inhibitor postpones methane hydrate formation. In other words, in this case maximums either do not shift or shift less than normal as time passes. Better kinetic inhibitors cause longer delays or no delay. As a result, RDF diagrams were calculated for each system (Fig. 6).

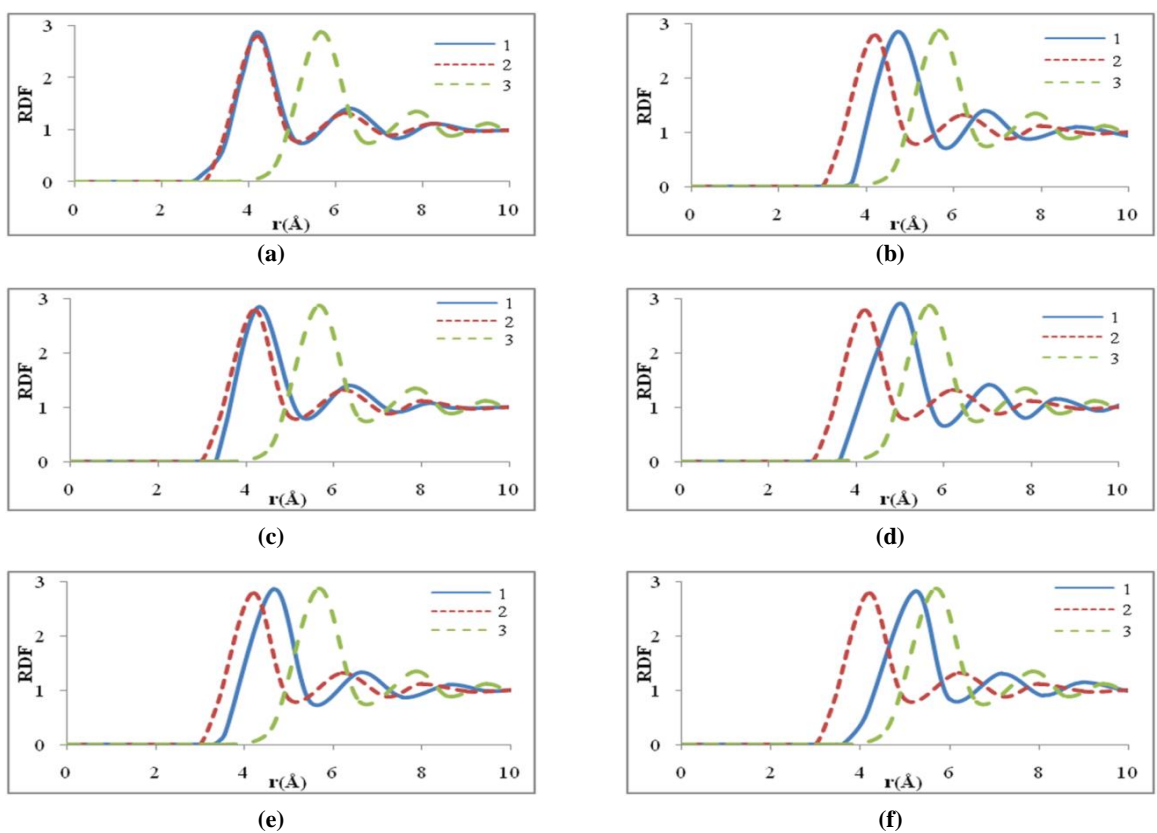

**Figure 6-** RDFs between carbon atoms ($g_{C-C}$) for systems containing ionic liquids; (1) $g_{C-C}$ for system containing ionic liquid after 35 ns, (2) $g_{C-C}$ for system at the beginning and (3) $g_{C-C}$ for blank system after 35 ns; (a) $[C_3(OH)_2mim][f_2N]$, (b) $[C_4mim][BF_4]$, (c) $[C_2OHmim][f_2N]$, (d) $[C_4mim][OAc]$, (e) $[C_2mim][BF_4]$ and (f) $[C_2mim][EtSO_4]$.

In system containing $[C_3(OH)_2mim][f_2N]$, maximum positions did not change, which means that $[C_3(OH)_2mim][f_2N]$ has the greatest kinetic inhibition effectiveness among others. Maximums in system containing $[C_2OHmim][f_2N]$ shifted a little, after 35 ns two maximums were located in 4.3 Å and 6.33 Å. Maximums in systems containing $[C_2mim][BF_4]$ and $[C_4mim][BF_4]$ appeared in 4.65 Å, 6.62 Å and 4.71 Å, 6.7 Å, respectively. Systems containing $[C_4mim][OAc]$ and $[C_2mim][EtSO_4]$ had the weakest kinetic inhibition effectiveness in hydrate formation, having maximums in 5.1 Å, 7.04 Å and 5.22 Å, 7.16 Å, respectively. Consequently, we can put kinetic inhibition effectiveness of ILs in methane hydrate formation in this order:

$[C_3(OH)_2mim][f_2N]>[C_2OHmim][f_2N]>[C_2mim][BF_4]>[C_4mim][BF_4]>[C_4mim][OAc]>[C_2mim][EtSO_4]$

## 4. Conclusion

The performance of six alkyl imidazolium ILs, as a class of gas hydrate inhibitors, has been investigated. Their thermodynamic and kinetic inhibition effects on methane hydrate crystal growth and methane hydrate formation were considered. These ILs were found to act as dual function hydrate inhibitors.

In order to understand the thermodynamic inhibition effectiveness, influence of ILs on structure of water phase was studied because the primary effect of the inhibitor is on the structure of the water phase. Systems containing $[C_3(OH)_2mim][f_2N]$ and $[C_2OHmim][f_2N]$ highly affect water phase. Thus, these two ILs are expected to be better thermodynamic inhibitors. Presence of OH functional groups in $[C_3(OH)_2mim]^+$ and $[C_2OHmim]^+$ is the reason for strong hydrogen bonds between these cations and water molecules and therefore great thermodynamic inhibition. $[C_3(OH)_2mim][f_2N]$ also seems to be a better thermodynamic inhibitor than methanol and NaCl, which are known as common thermodynamic inhibitors.

The kinetic inhibition effectiveness of ILs was also investigated in two parts. At first, the kinetic inhibition of methane hydrate growth was analyzed. Simulation results showed that the effect of cation type on the kinetic inhibition effectiveness is more significant than that of anion type. Cations tended to move towards the hydrate crystal and bond to hydrate surface and prevent water molecules from gathering close to hydrate surface.

In the second part, kinetic inhibition of methane hydrate formation was studied. For this purpose, long running simulations were carried out. Although methane hydrates could not be observed to be formed after 35 ns, RDFs enable us to follow methane hydrate formation. Results showed that $[C_3(OH)_2mim][f_2N]$ postponed methane hydrate formation more than other ILs.

In summary, among the ILs studied here $[C_3(OH)_2mim][f_2N]$ can offer greater thermodynamic inhibition effects than the conventional thermodynamic inhibitors. However, at the same time, all the ILs also delay hydrate formation, an advantage that the conventional thermodynamic inhibitors do not show. Numerous combinations of cations and anions that can be tailored to form ILs enable us to find more effective gas hydrate inhibitors that are inexpensive, available and biodegradable.

*ACKNOWLEDGMENT*

This research was sponsored by the Research and Development Department of National Iranian Offshore Oil Company (NIOOC). The authors are thankful to Professor Ponder for the TINKER software package.